\documentclass[11pt]{article}
\pdfoutput=1
\usepackage{jheppub}
\usepackage{bm, bbm, slashed}
\usepackage{subcaption}

\newcommand{\beq}{\begin{eqnarray}}
\newcommand{\eeq}{\end{eqnarray}}

\title{Absence of Power-Law Mid-Infrared Conductivity in Gravitational Crystals}

\author{Brandon W. Langley,}
\author{Garrett Vanacore,}
\author[1]{Philip W. Phillips
\note{Guggenheim Fellow}}
\affiliation{Department of Physics and Institute for Condensed Matter Theory \\ University of Illinois, 1110 W. Green Street, Urbana, IL 61801}

\emailAdd{blangle2@illinois.edu}
\emailAdd{vanacor2@illinois.edu}
\emailAdd{dimer@illinois.edu}


\arxivnumber{1506.06769}

\abstract{We compute conductivities of strongly-interacting and non-uniform charge densities dual to inhomogeneous anti-de Sitter--black hole spacetimes. Backreacting bulk scalars with periodic boundary profiles, we construct generalizations of Reissner-Nordstr\"om-AdS that interpolate between those used in two previous studies --- one that reports power-law scaling for the boundary optical conductivity and one that does not. We find no evidence for power-law scaling of the conductivity, thereby corroborating the previous negative result that gravitational crystals are insufficient to generate the power-law mid-infrared conductivity observed in cuprate superconductors.
}

\notoc

\begin{document}
\maketitle
\flushbottom

\paragraph{}
Power laws are ubiquitous to critical phenomena as they are the fingerprint of scale invariant correlations.  To no surprise then, quantum criticality is the most commonly proffered explanation\cite{anderson,ogata1,mfl} for the power-law scaling
\beq
|\sigma(\omega)|\sim\omega^{-2/3}
\label{midir}
\eeq
observed in the mid-infrared frequency range of the optical conductivity\cite{marel,ElAzrak1994,Hwang2007,Schlesinger1990} for optimally doped copper-oxide superconductors.   However, we currently have no microscopic understanding of how quantum criticality emerges from the strong correlations that mediate the normal state near optimal doping.   Such an understanding requires precise knowledge of the low-energy degrees of freedom in the strongly coupled regime.  Ascertaining these degrees of freedom has proven difficult because the integral of the optical conductivity\cite{uchida,basovopt1} up to the optical gap exceeds the number of doped holes.  Consequently, no one-to-one mapping\cite{phillipsrmp} exists between the number of doped holes and the actual propagating degrees of freedom at low energy.

Given the seemingly unbridgeable divide between the ultraviolet (UV) and infrared (IR) physics in the cuprates, it is desirable to address the problem with a method which is based on conserved currents rather than on a traditional particle description. The gauge/gravity duality offers such an alternative, in which a strongly-coupled, conformal, $d$-dimensional quantum field theory lives at the boundary of a $(d+1)$-dimensional gravity theory. The workhorse in condensed matter applications is Reissner-Nordstr\"om-AdS (RN-AdS), which provides the simplest bulk description of matter at finite charge density. RN-AdS is often treated as a static background to be augmented by a set of probe fields which source additional operators at the boundary. Horowitz, Santos, and Tong (HST)\cite{jorge} improved upon these treatments by solving the completely coupled equations of motion for gravity, electromagnetism, and a neutral scalar field. Further, they fixed the boundary source provided by the scalar to be $ A_0\cos kx$, thereby imprinting a periodic distortion upon the bulk geometry and the boundary charge density. Taking this inhomogeneous bulk as a new static background, the boundary conductivity was then obtained by solving equations of motion for propagating fluctuations of the bulk fields. The key finding of reference \cite{jorge} is that the conductivity obtains --- in addition to a Drude peak --- a mid-infrared power law identical to that seen in the cuprates, eq.\ \eqref{midir}, plus an additive constant.  HST obtained identical power laws for an ionic lattice --- charged matter with a spatially periodic chemical potential, $\lim_{z\to 0}A_t=\mu(x)$ --- in a slightly different geometry\cite{santos2}. These conclusions are indeed startling and imply that a gravitational crystal encodes the optical conductivity of the cuprates.  

However, the key claim that Einsteinian gravity, a Maxwell field and an inhomogeneous charge density encode the mid-infrared conductivity of the cuprates has been called into question\cite{donos,Rangamani2015}.   In reference \cite{donos} Donos and Gauntlett (DG) studied a model inspired by the Q-ball potential of Coleman\cite{coleman} which has the added simplification that only ordinary rather than partial differential equations need be solved to obtain the conductivity. Their scalar field, of the form $\Phi(z,x)=\phi(z)e^{ikx}$, leads to a uniform charge density as they chose a potential of the form $V(|\Phi|^2)$. Added differences with the work of HST is the use of a scalar mass of $m^2=-3/2$\footnote{Extremal RN-AdS$_4$ has an emergent $AdS_2\times \mathbb{R}^2$ geometry near the horizon, which hosts a quantum dual with a Breitenlohner-Freedman (BF) bound higher than that of the boundary theory. Donos et al.\ argue that the HST results may be unstable because their chosen scalar field mass violates the BF bound in the near-horizon CFT.} and the radial gauge as opposed to the de Donder gauge and Lorenz gauges.  In addition, rather than using a log--log plot to discern the presence of a power law, they plotted
\beq
\alpha=1+\omega\frac{|\sigma|''}{|\sigma|'}.
\eeq
A value of $\alpha=-2/3$ would correspond to the mid-infrared power-law conductivity of the cuprates.  None was found.  In fact, $\alpha$ was found to vary fairly widely as a function of frequency precisely where the power law was reported by HST.  Similar results from a slightly different model have also been reported in reference \cite{Rangamani2015}.

Given the substantial differences between the HST and DG constructions, it is instructive to resolve this problem through a new construction that interpolates between the two models. We achieve this by introducing two neutral scalars, subject to periodic potentials with tunable phase angles, to four-dimensional RN-AdS. Although the HST and DG setups differ drastically in terms of the boundary charge density, we find that for identical system parameters used by HST,  $\alpha$ deviates significantly from the constant needed to reproduce the mid-infrared power law of the cuprate conductivity.  Consequently, we conclude that Einsteinian gravitational crystals are insufficient to explain the power-law scaling in the cuprates.

We consider the Einstein-Maxwell action
\beq \label{eq:RN}
		S = \frac{1}{16\pi G_N}\int d^4 x \sqrt{-g}\left(R-2\Lambda-\frac{1}{2}F^2\right),
\eeq
where $\Lambda=-6/L^2$ is the cosmological constant, $L$ is the AdS length, $F=dA$ is the field strength of a Maxwell field, $R$ is the Ricci scalar, $g$ is the metric determinant, and $G_N$ is the Newton constant. To this we append an action for two neutral scalar fields,
\begin{gather}
	S_{\Phi} = \frac{1}{16 \pi G_N} \int d^4 x  \sqrt{-g} 		 \sum_{i=1}^2\left[2\left(\nabla\Phi_i\right)^2+4V\left(\Phi_i\right)\right], \nonumber \\
	V(\Phi) = \frac{m^2}{2L^2}\Phi^2.
\end{gather}
The asymptotic expansion for the scalar fields is
\begin{gather} \label{eq:twoscalar}
		 \Phi_i=z^{3-\Delta}\Phi_i^{(1)}+z^\Delta\Phi_i^{(2)}+\cdots, \nonumber\\ 
		\Phi_1^{(1)}(x) = A_1\cos\left(k_1x-\frac{\theta}{2}\right), \nonumber \\ 
		\Phi_2^{(1)}(x) = A_2\cos\left(k_2x+\frac{\theta}{2}\right),
\end{gather}
where $\Delta=3/2+\sqrt{9/4+m^2}$. 

Obtaining the boundary conductivity requires two distinct calculations. First, solve the full equations of motion for a static background.  Second, kick the system with a time-dependent electric field and find the current response.  As all of this is standard (see references \cite{headrick,jorge,santos2,donos2}), we will only highlight the key features.  To assist in solving the equations of motion,
\begin{align} \label{eq:EOM}
	&E_{ab} \equiv R_{ab} - \Lambda g_{ab} - \hat{T}_{ab} = 0, \\
	&\nabla^a F_{ab} = 0, \\
	&\Box \Phi_i - V'\left(\Phi_i\right) = 0,
\end{align}
we solved instead the  Einstein-DeTurck equation
\begin{gather}
\label{eq:DeTurck}
		E_{ab} = \nabla_{(a}\xi_{b)}, \\
\label{eq:xi}
		 \xi^a \equiv g^{cd}\left(\Gamma^a_{cd}(g)-\Gamma^a_{cd}(\overline{g})\right).
\end{gather}
Here $\hat{T}_{ab}\equiv T_{ab} - \left(g_{ab}/2\right) T$ is the trace-reversed energy-momentum tensor and $\overline{g}$ is a reference metric. To directly compare with the HST results, we use the mass value $m^2 = -2$. The Einstein-DeTurck equation simultaneously encodes $E_{ab}=0$ and the harmonic gauge $\xi^a=0$ due to $E_{ab}$ obeying a continuity equation. Because the DeTurck term breaks gauge invariance, metric components can be turned on to compensate any terms from $\hat{T}_{ab}$. We take the ansatz
\begin{gather}
ds^2 = \frac{L^2}{z^2} \Big[ -(1-z) P(z) Q_{tt} dt^2 + \frac{Q_{zz} dz^2}{(1-z)P(z)}\nonumber\\ \qquad+ Q_{xx} (dx + z^2 Q_{zx} dz)^2 + Q_{yy}dy^2 \Big], \nonumber\\
		P(z) = 1+z+z^2-\frac{\mu_1^2}{2}z^3, \nonumber\\
	A = (1-z)a_t(z,x)dt,\quad \Phi_i = z\phi_i (z,x).
\end{gather}
Here all coordinates have been rescaled to be dimensionless and the radial coordinate $z$ is parameterized to extend from $[0,1]$. The conformal boundary exists at $z=0$ and an event horizon exists at $z=1$. The temperature is given by $T=P(1)/(4\pi L)$. This ansatz clearly reduces to Reissner-Nordstr\"om-AdS (RN-AdS) when $Q_{tt}=Q_{zz}=Q_{xx}=Q_{yy}=1, Q_{zx}=0$, $a_t=\mu_1=\mu$ and $\phi_i=0$, which is used as our reference metric $\overline{g}$ in (\ref{eq:xi}).

If we set $A_1=A_2=A_0$ and $k_1=k_2=k$, then the parameter $\theta$ tunes this model between the HST and Q-lattice models. $\theta=0$ corresponds to the HST lattice and $\theta=\pi/2$ yields the Q-lattice. Consequently, regardless of the origin of the power law, the model considered here should be able to unearth the source.

We have eight equations of motion to solve. At the conformal boundary, $z=0$, we imposed Dirichlet boundary conditions.  These conditions will produce the desired AdS geometry, fix the chemical potential, and fix the scalar lattice. The remaining conditions will be so-called regularity conditions. We demand that all the functions can be expanded at the horizon $z=1$ as
	\begin{align} \label{eq:regularity}
		& Q_{ab}(z,x) = Q_{ab}^{(0)}(x)+(1-z)Q_{ab}^{(1)}(x)+\cdots, \nonumber\\
		& a_t(z,x) = a_t^{(0)}(x)+(1-z)a_t^{(1)}(x)+\cdots, \nonumber\\
		& \phi_i(z,x) = \phi_i^{(0)}(x)+(1-z)\phi_i^{(1)}(x)+\cdots.
	\end{align}
We then plug these expansions into various equations. Two further conditions can be specified by setting $\xi^z=\xi^x=0$ and the remaining six will be set by demanding that the equations of motion be satisfied to lowest order in the expansion. This will fix relationships among the $(0)$ and $(1)$ coefficients, giving Robin type boundary conditions at $z=1$.
The numerical solution was obtained through the Newton-Raphson method using a sixth-order finite difference grid for the $z$-direction and a Fourier grid for the $x$-direction.  Details of such methods can be found in reference \cite{trefethen}.

\begin{figure}[t]
		\centering
		\includegraphics[scale=.98]{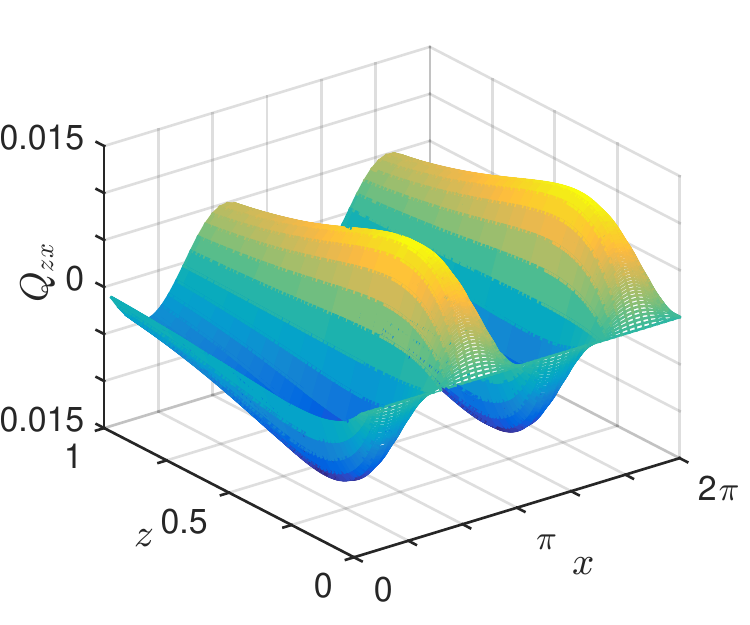}
		\includegraphics[scale=.98]{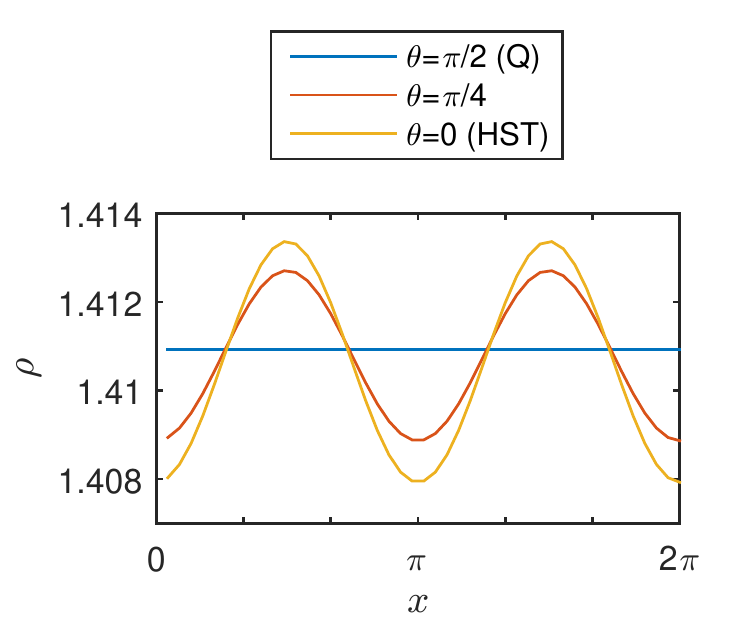}
		\captionsetup{justification=raggedright,singlelinecheck=false}
		\caption{(Left) Plot of $Q_{zx}$ for $A_0=0.75/\sqrt{2}$, $k=1$, $\theta=0$, $\mu=1.4$, and $T/\mu=0.115$. For this plot $\max\left(|\xi^a|\right) \lesssim 10^{-12}$. (Right) Plot of the charge density $\rho=\lim_{z\to 0}\sqrt{-g}F^{tz}$ for same parameters, and with $\theta = \pi/4, \pi/2$.} \label{fig:static}
\end{figure}

In figure \ref{fig:static} we have a sample plot of the off-diagonal component $Q_{zx}$. It is clear that the periodicity of the scalar field is imprinted on the background (in multiples of $2k$ due to the quadratic form of the scalar field in the energy-momentum tensor). Figure \ref{fig:static} also contains the boundary charge density for different values of $\theta$, distinguishing the periodic imprint of the HST lattice versus the uniformity of the Q-lattice. Figure \ref{fig:static} was computed in MATLAB at double precision with (300,45) grid points on the $(z,x)$ plane, which is a typical grid size for the calculations in this paper.

The second step involves a time-dependent perturbation around the static solution which enables a calculation of the conductivity. We follow the methods of reference \cite{donos2}. Denoting background fields with bars, we write
\beq
		g_{ab}=\overline{g}_{ab}+h_{ab},\;\; A_a=\overline{A}_a+b_a,\;\; \Phi_i=\overline{\Phi}_i+\eta_i,
\eeq
where the barred quantities are the static background and the extra pieces $h_{ab}$, $b_a$, and $\eta_i$ are fluctuations. The leading term in $b_x$ sources an electric field at the boundary and the subleading term contains the current response.  The Kubo formula can then be used to obtain the conductivity. 

\begin{figure}[t!]
	\begin{subfigure}[t]{0.5\textwidth}
		\centering
		\includegraphics[scale=1]{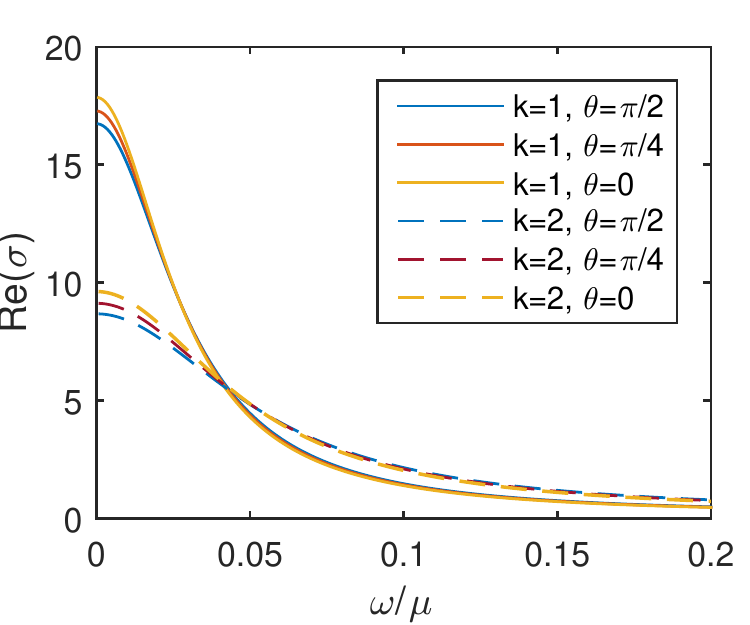}
	\end{subfigure}
	\begin{subfigure}[t]{0.5\textwidth}
		\centering
		\includegraphics[scale=1]{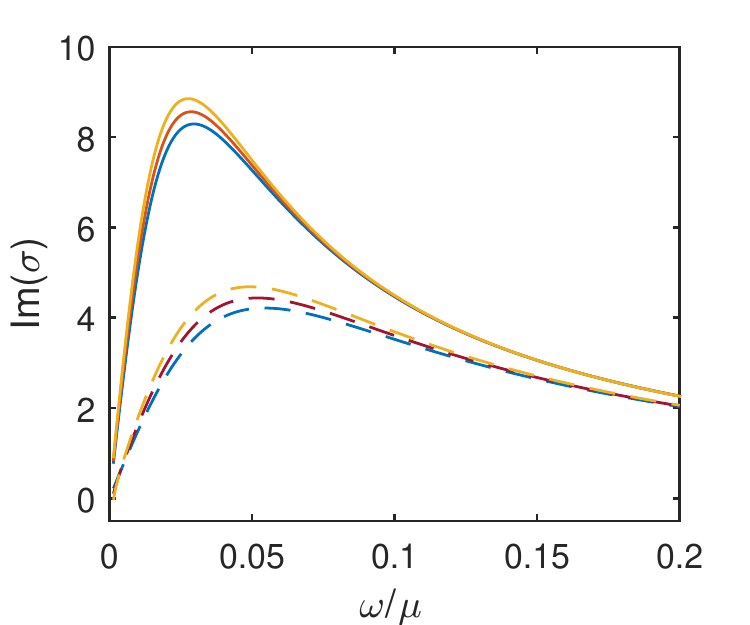}
	\end{subfigure}

		\captionsetup{justification=raggedright,singlelinecheck=false}
		\caption{Low frequenecy plots of real and imaginary parts of the conductivity for various parameters. In each plot $A_0/k=3/(4\sqrt{2})$, $\mu=1.4$ and $T/\mu=0.115$.}
		\label{RIS}
\end{figure}
\begin{figure}[t!]
	\begin{subfigure}[t]{0.5\textwidth}
		\centering
		\includegraphics[scale=1]{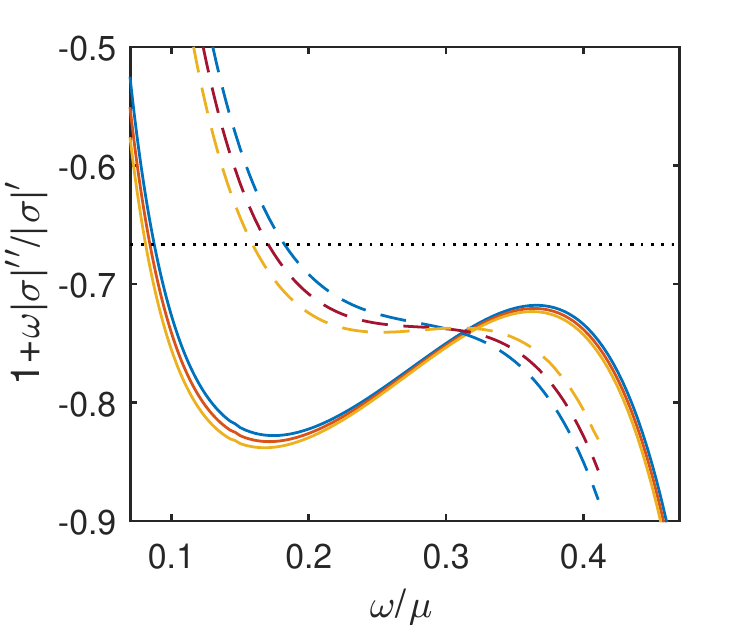}
	\end{subfigure}
	\begin{subfigure}[t]{0.5\textwidth}
		\centering
		\includegraphics[scale=1]{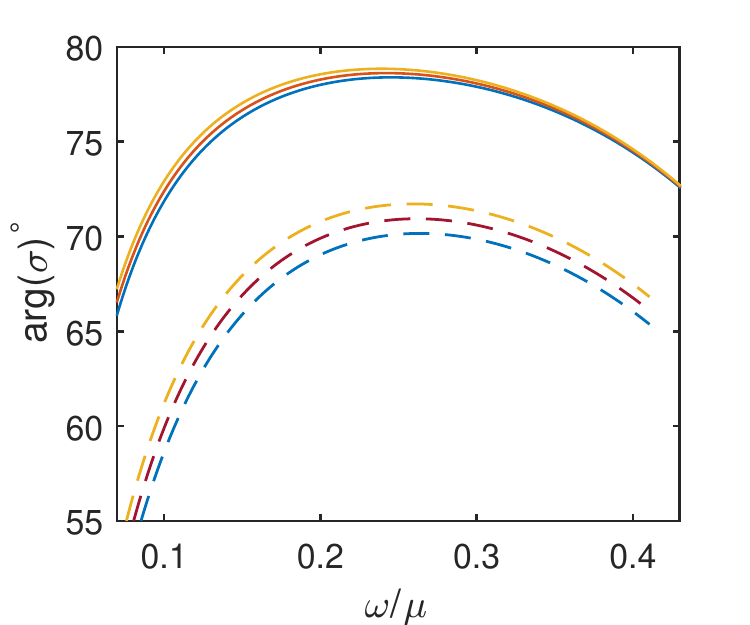}
	\end{subfigure}

		\captionsetup{justification=raggedright,singlelinecheck=false}
		\caption{Plots of the power function and the argument of the conductivity for the same parameters as figure\ \ref{RIS}. The dotted line marks $-2/3$. Refer to figure \ref{RIS} for legend.}
		\label{powerlaw}
\end{figure}

Using this method, we computed the conductivity as a function of the interpolating parameter $\theta$.  Shown in figure \ref{RIS} are the real and the imaginary parts of the conductivity for two different values of $k$ and three values of $\theta$. For each choice of parameters the low frequency conductivity obeys the Drude form. Note, even though $\theta=\pi/2$ corresponds to a uniform charge density of DG, the conductivities are almost identical to those of HST ($\theta=0$). 

\begin{figure}
	\centering
	\includegraphics[scale=1]{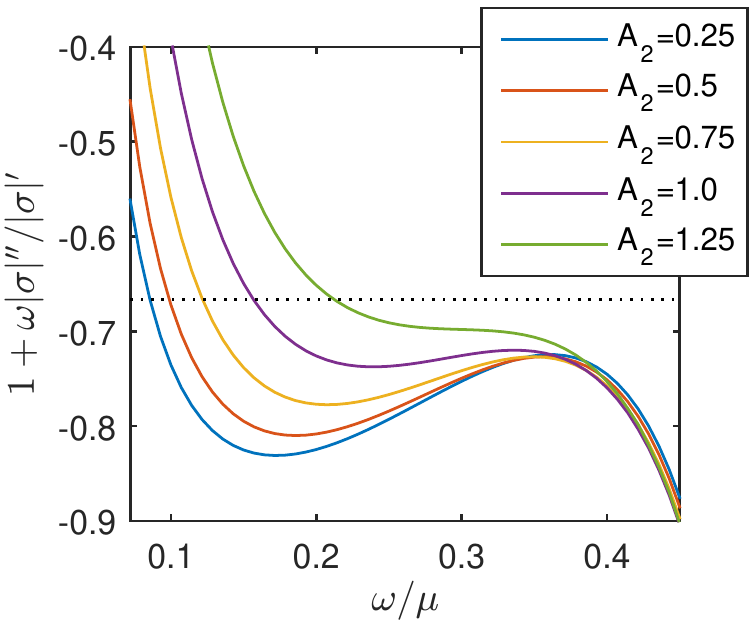}
	\captionsetup{justification=raggedright,singlelinecheck=false}
	\caption{This plot was generated using $A_1=0.75$, $k_1=1$, $k_2=2$, $\theta=0$, $\mu=1.4$ and $T/\mu=0.115$. The amplitude $A_2$ of the higher harmonic lattice was adjusted.}
	\label{HH}
\end{figure}

Figure \ref{powerlaw} is the key test for the existence of the power-law conductivity. In the left panel we show a couple of plots of $\alpha$ using the parameter choices of reference \cite{jorge}. The dotted line indicates a power law of $-2/3$.  As is evident, regardless of the value of $\theta$, no discernible power law exists even as $k$ is varied. Also of note is the fact that the DG ($\theta=\pi/2$) and HST  ($\theta=0$) models yield almost identical numerical results for the conductivity.  The right panel of figure \ref{powerlaw} presents the phase angle which also deviates from $60^\circ$.  

As a final check we also varied the amplitude $A_2$ for a dichromatic lattice in figure \ref{HH}, as this could introduce a mix of higher harmonics.  In this case as well, no evident power law exists.  In fact, for any range of parameters including temperature, no power law was found.

Since we have introduced a model that is capable of interpolating between DG and HST and we find no power law in either case, we conclude that gravitational crystals, although adequate in describing Drude response,  do not encode the power-law optical conductivity of the cuprates. 

Regarding the origin of the power-law optical conductivity, the only study\cite{limtragool} to date that has successfully reproduced the $\omega^{-2/3}$ scaling relies on excitations which exist on all energy scales --- namely scale-invariant matter or unparticles.  Given that the radial direction in AdS represents the running of coupling constants, in principle it contains the correct ingredients to capture unparticle excitations.  Hence, we anticipate that some construction using gauge/gravity duality, other than the one presented here, should be able to reproduce the power law.

\acknowledgments
We thank Aristos Donos and Jorge Santos for their generosity in assisting us with the numerics.  P. Phillips thanks the NSF DMR-
1104909 and the J. S. Guggenheim Foundation for financial support. 
\bibliographystyle{JHEP}
\bibliography{optcond-1}

\end{document}